\RequirePackage{fix-cm}
\documentclass[smallextended]{svjour3}       
\smartqed  
\usepackage{graphicx}

\usepackage{amssymb}
\usepackage{amsmath}
\usepackage{bm}
\usepackage{mathptmx}

\begin{document}

\title{Gravitational mass and energy gradient in the ultra-strong magnetic fields
}


\author{Zi-Hua Weng
}


\institute{Zi-Hua Weng \at
              School of Physics and Mechanical \& Electrical Engineering, Xiamen University, Xiamen 361005, China    \\
              \email{xmuwzh@xmu.edu.cn}           
}

\date{Received: date / Accepted: date}

\maketitle

\begin{abstract}
  The paper aims to apply the complex octonion to explore the influence of the energy gradient on the E\"{o}tv\"{o}s experiment, impacting the gravitational mass in the ultra-strong magnetic fields. Until now the E\"{o}tv\"{o}s experiment has never been validated under the ultra-strong magnetic field. It is aggravating the existing serious qualms about the E\"{o}tv\"{o}s experiment. According to the electromagnetic and gravitational theory described with the complex octonions, the ultra-strong magnetic field must result in a tiny variation of the gravitational mass. The magnetic field with the gradient distribution will generate the energy gradient. These influencing factors will exert an influence on the state of equilibrium in the E\"{o}tv\"{o}s experiment. That is, the gravitational mass will depart from the inertial mass to a certain extent, in the ultra-strong magnetic fields. Only under exceptional circumstances, especially in the case of the weak field strength, the gravitational mass may be equal to the inertial mass approximately. The paper appeals intensely to validate the E\"{o}tv\"{o}s experiment in the ultra-strong electromagnetic strengths. It is predicted that the physical property of gravitational mass will be distinct from that of inertial mass.

\keywords{gravitational mass \and inertial mass \and electromagnetic strength \and quaternion \and octonion}

\PACS{04.50.-h \and 04.80.Cc \and 11.10.Kk \and 02.10.De \and 04.20.Cv}

 \subclass{83E15 \and 83C65 \and 17A35 \and 35Q75}

\end{abstract}

\section{Introduction}

The validity of equivalence principle, in the General Theory of Relativity (GR for short), has been intriguing and puzzling the scholars since a long time. Some scholars are suspicious of the equivalence principle \cite{weng1}, validating the relevant issues from various aspects, developing a few new theories to violate the postulate \cite{damour}. The suspicion keeps coming until now. Especially, the E\"{o}tv\"{o}s experiment has never been validated under any ultra-strong gravitational or electromagnetic field. The postulate has been becoming a vulnerable point to be criticized. It seems that this postulate may be inappropriate, and even unnecessary. Until recently, the appearance of the electromagnetic and gravitational theories, described with the complex-octonions, replies to a part of the suspicion. In this complex-octonion field theory, a few properties of gravitational mass are able to be predicted, including the variation, influencing factors, and energy gradient (a force term) and so forth.

In the 20th century, some scholars applied a variety of materials to validate experimentally the E\"{o}tv\"{o}s experiment. Up to now the measurement accuracy of E\"{o}tv\"{o}s experiment achieved the order of $10^{-17}$ in the gravitational field. L. E\"{o}tv\"{o}s measured the influence of the earth's rotation on the inertial and gravitational mass. Subsequently, Dicke \emph{et al.} \cite{dicke} measured the influence of the Sun on the inertial and gravitational mass. Braginsky \cite{braginsky} modified the experimental instrument similar to that of Dicke, and improved the measurement accuracy. Bondi \cite{bondi} pointed out that there might be the differentiation between the active gravitational mass and the passive gravitational mass. And it means that the E\"{o}tv\"{o}s experiment is only about the experimental validation between the passive gravitational mass and the inertial mass. Beall analyzed the E\"{o}tv\"{o}s experiment about the elementary particle \cite{beall}, and doubted the equivalence of inertial and gravitational mass. Owing to the extremely experimental difficulty, the study is still rest on the theoretical analysis at the present time. Olson \emph{et al.} \cite{olson} explored the active gravitational mass of the moving object, by means of the collision among the particles. The method may exist a few flaws, moreover it has to be dealt with the problem of the transverse and longitudinal mass.

Subsequently Fischbach \cite{fischbach} suspected the outcome of E\"{o}tv\"{o}s experiment, and noticed that there was one distinct discrepancy between the gravitational constant measured in the laboratory with that observed in the geophysics. And then he proposed that there might be one kind of non-Newtonian coupling term between two objects. Niebauer \emph{et al.} \cite{niebauer} carried out a direct free-fall experiment in the Earth's gravitational field, in order to measure the differential acceleration between two different materials. And it put the limit on the strength and range of the proposed fifth-force. Stubbs \emph{et al.} \cite{stubbs} placed a 1-ton Pb source close to a Be/Al torsion balance, to search the fifth-force. Adelberger \emph{et al.} \cite{adelberger} observed the differential acceleration of two different test body pairs, to search for feeble, macroscopic forces arising from the exchange of hypothetical ultra-low-mass bosons. Shapiro \emph{et al.} \cite{shapiro} applied the echo delay of laser signals transmitted from the Earth and reflected from cube corners on the Moon, to measure the gravitational binding energy to contribute equally to Earth's inertial and passive gravitational masses. Baessler \emph{et al.} \cite{baessler} modified the lunar-ranging test of the equivalence principle for gravitational self-energy, to compare the accelerations of miniature Earth and Moon toward the Sun. Reasenberg \emph{et al.} \cite{reasenberg} developed a Galilean test of the equivalence principle, in which two pairs of test mass assemblies were in free fall in a vacuum chamber. Schlamminger \emph{et al.} \cite{schlamminger} used a continuously rotating torsion balance instrument, to measure the acceleration difference of beryllium and titanium test bodies. Moffat \cite{moffat} suggested that applying the E\"{o}tv\"{o}s experiment in the satellite to test the weak equivalence principle for zero-point vacuum energy.

But success in recent years also produced disappointment. Up to now most of E\"{o}tv\"{o}s experiments are only validated experimentally in the quite weak gravitational field \cite{weng2} . It means that the previous deduction of E\"{o}tv\"{o}s experiment is only fit for the case of the weak gravitational field. Further it is insufficient to claim this deduction to be suitable to other circumstances. Especially the E\"{o}tv\"{o}s experiment has never been verified in the strong magnetic field. It is aggravating the perplexing qualm about disputing the deduction of E\"{o}tv\"{o}s experiment.

Maxwell \cite{maxwell} was the first to apply the algebra of quaternions to describe the electromagnetic theory. Nowadays some scholars applied the complex quaternion \cite{weng3} and complex octonion \cite{weng4} to describe the electromagnetic and gravitational fields. Morita \cite{morita} studied the quaternion field theory. Anastassiu \emph{et al.} \cite{anastassiu} applied the algebra of quaternions to depict the electromagnetic feature. Doria \cite{doria} utilized the quaternion to express the gravitational theory. Majernik \cite{majernik} extended the Maxwell-like gravitational field equations making use of the quaternion. Edmonds \cite{edmonds} adopted the quaternion to elucidate the gravitational theory in the curved space-time. Meanwhile Rawat \emph{et al.} \cite{rawat} discussed the gravitational field equation with the quaternion treatment. A few scholars depicted the electromagnetic and gravitational field with the complex octonion. Mironov \emph{et al.} \cite{mironov} depicted the electromagnetic equations with the algebra of octonions. Demir \emph{et al.} \cite{demir2} applied the octonion to research the gravitational field equations. Gogberashvili \cite{gogberashvili} explored the electromagnetic theory with the octonion. Negi \emph{et al.} \cite{negi2} described the Maxwell's equations by means of the octonion. Furui \cite{furui} discussed the axial anomaly and the triality symmetry of octonions, making use of the quaternions in the Yang-Mills Lagrangian.

The paper adopts the complex octonion to depict the electromagnetic and gravitational fields simultaneously. According to the field theory, the gravitational mass relates to the gravitational and electromagnetic strengths. The comparatively strong electromagnetic strength will result in the observable variation of gravitational mass. Meanwhile the field theory predicts one new kind of energy gradient, that is, one new force term. This force term is independent of either the electric charge and mass, or the direction of electromagnetic strength. But this force term relates with the gradient of electromagnetic strength, and may impact the previous state of equilibrium in the E\"{o}tv\"{o}s experiment. And it predicts that the existing inferences of E\"{o}tv\"{o}s experiment will be violated in the strong electromagnetic field.

Only if the equivalence principle, which is relevant to the equivalence of inertial mass and gravitational mass, is introduced to become one of two postulates, the scholars can construct the GR. Presenting a striking contrast to the above is that it is still able to establish the gravitational and electromagnetic theories in the complex-octonion curved-spaces (see Ref.[1]), even though we take no account of the equivalence principle in the recent studies. Further it is capable of deducing some formulae between the physical quantity and spatial parameter \cite{weng5}, in the curved composite-space (a function space) described with the complex octonions. The GR and recent study both succeed to the Cartesian academic thought of `the space is the extension of substance (or even field)'. In other words, no matter whether the gravitational mass is variable or not, it is still able to construct the gravitational and electromagnetic theories in the curved spaces. And that the GR can be considered as one of special cases only.

During the founding period of GR, the result of E\"{o}tv\"{o}s experiment might elicit certain intuitions or inspirations to help A. Einstein, heightening his confidence on the faith there must be the curved space in the physics, acting as an important part to the GR. However, the intuition and inspiration, which are similar to the temporary scaffolding in the process of building, are not always accurate or even necessary. And it is possible that the temporary scaffolding may need to be updated, in the succeeding building maintenance. Subsequently, for the field theories in the curved space, the E\"{o}tv\"{o}s experiment is becoming increasingly isolated, and its influence power is declining day by day. No matter whether the E\"{o}tv\"{o}s experiment is validated or not, it should not become a crucial hindrance or judgment, for the establishment of field theory in the curved space. On the basis of these recognitions, all of us are able to explore the E\"{o}tv\"{o}s experiment comparatively objectively and neutrally, under some extreme conditions, for instance, the ultra-strong magnetic fields.

\section{Field equations}

The octonion space $\mathbb{O}$ can be separated into two orthogonal quaternion spaces, $\mathbb{H}_g$ and $\mathbb{H}_e$ , that is, $\mathbb{O} = \mathbb{H}_g + \mathbb{H}_e$. The quaternion space $\mathbb{H}_g$ can be applied to describe the feature of gravitational field, while the $S$-quaternion space $\mathbb{H}_e$ is able to be used for depicting the property of electromagnetic field.

In the quaternion space $\mathbb{H}_g$ , the basis vector is $\emph{\textbf{i}}_j$ , the radius vector is $\mathbb{R}_g = i r_0 \emph{\textbf{i}}_0 + \Sigma r_k \emph{\textbf{i}}_k$ , and the velocity is $\mathbb{V}_g = i v_0 \emph{\textbf{i}}_0 + \Sigma v_k \emph{\textbf{i}}_k$ . The gravitational potential is $\mathbb{A}_g = i a_0 \emph{\textbf{i}}_0 + \Sigma a_k \emph{\textbf{i}}_k$ , the gravitational strength is $\mathbb{F}_g = f_0 \emph{\textbf{i}}_0 + \Sigma f_k \emph{\textbf{i}}_k$, and the gravitational source is $\mathbb{S}_g = i s_0 \emph{\textbf{i}}_0 + \Sigma s_k \emph{\textbf{i}}_k$ . Herein $r_j$ , $v_j$ , $a_j$, $s_j$ , and $f_0$ are all real. $f_k$ is one complex number. $i$ is the imaginary unit. $\emph{\textbf{i}}_0 = 1$, $\emph{\textbf{i}}_k^{~2} = -1$. $j = 0, 1, 2, 3$. $k = 1, 2, 3$.

In the $S$-quaternion space $\mathbb{H}_e$ , the basis vector is $\emph{\textbf{I}}_j$ , the radius vector is $\mathbb{R}_e = i R_0 \emph{\textbf{I}}_0 + \Sigma R_k \emph{\textbf{I}}_k$, and the velocity is $\mathbb{V}_e = i V_0 \emph{\textbf{I}}_0 + \Sigma V_k \emph{\textbf{I}}_k$. The electromagnetic potential is $\mathbb{A}_e = i A_0 \emph{\textbf{I}}_0 + \Sigma A_k \emph{\textbf{I}}_k$, the electromagnetic strength is $\mathbb{F}_e = F_0 \emph{\textbf{I}}_0 + \Sigma F_k \emph{\textbf{I}}_k$, and the electromagnetic source is $\mathbb{S}_e = i S_0 \emph{\textbf{I}}_0 + \Sigma S_k \emph{\textbf{I}}_k$. Herein $\mathbb{H}_e = \mathbb{H}_g \circ \emph{\textbf{I}}_0$. The symbol $\circ$ denotes the octonion multiplication. $R_j$, $V_j$, $A_j$, $S_j$, and $F_0$ are all real. $F_k$ is one complex number. $\emph{\textbf{I}}_j^{~2} = -1$.

In the complex octonion space $\mathbb{O}$ , the octonion radius vector is $\mathbb{R} = \mathbb{R}_g + k_{eg} \mathbb{R}_e$, the octonion velocity is $\mathbb{V} = \mathbb{V}_g + k_{eg} \mathbb{V}_e$, with $k_{eg}$ being a coefficient. The octonion field potential is $\mathbb{A} = \mathbb{A}_g + k_{eg} \mathbb{A}_e$ , the octonion field strength is $\mathbb{F} = \mathbb{F}_g + k_{eg} \mathbb{F}_e$. Apparently $\mathbb{V}$, $\mathbb{A}$, $\mathbb{F}$, and $\mathbb{S}$ and so forth are all octonion functions of $\mathbb{R}$ .

The octonion definition of field strength is,
\begin{eqnarray}
\mathbb{F} = \lozenge \circ \mathbb{A} ~,
\end{eqnarray}
where $\mathbb{F}_g = \lozenge \circ \mathbb{A}_g$, and $\mathbb{F}_e = \lozenge \circ \mathbb{A}_e$. The quaternion operator is $\lozenge = i \emph{\textbf{i}}_0 \partial_0 + \Sigma \emph{\textbf{i}}_k \partial_k$. $\nabla = \Sigma \emph{\textbf{i}}_k \partial_k$, $\partial_j = \partial / \partial r_j$. $r_0 = v_0 t$. $v_0$ is the speed of light, and $t$ is the time.

In the quaternion space $\mathbb{H}_g$ , the gauge condition is, $- f_0 = \partial_0 a_0 - \nabla \cdot \textbf{a} = 0$. So $\mathbb{F}_g$ is reduced to $\textbf{f} = i \textbf{g} / v_0 + \textbf{b}$ . The gravitational acceleration is, $\textbf{g} / v_0 = \partial_0 \textbf{a} + \nabla a_0$ , and the other component of gravitational strength is $\textbf{b} = \nabla \times \textbf{a}$, which is similar to the magnetic flux density (Table 1). In the $S$-quaternion space $\mathbb{H}_e$ , the gauge condition is, $- \textbf{F}_0 = \partial_0 \textbf{A}_0 - \nabla \cdot \textbf{A} = 0$. Therefore $\mathbb{F}_e$ is reduced to $\textbf{F} = i \textbf{E} / v_0 + \textbf{B}$. The electric field intensity is $\textbf{E} / v_0 = \partial_0 \textbf{A} + \nabla \circ \textbf{A}_0$ , meanwhile the magnetic flux density is $\textbf{B} = \nabla \times \textbf{A}$. Herein $\textbf{a} = \Sigma a_k \emph{\textbf{i}}_k$ , $\textbf{f} = \Sigma f_k \emph{\textbf{i}}_k$ . $\textbf{A}_0 = A_0 \emph{\textbf{I}}_0$ , $\textbf{A} = \Sigma A_k \emph{\textbf{I}}_k$ . $\textbf{F}_0 = F_0 \emph{\textbf{I}}_0$ , $\textbf{F} = \Sigma F_k \emph{\textbf{I}}_k$ .

In the electromagnetic and gravitational fields, the octonion field source $\mathbb{S}$ is defined as  (see Ref.[20]),
\begin{eqnarray}
\mu \mathbb{S} && = - ( i \mathbb{F} / v_0 + \lozenge )^\ast \circ \mathbb{F}  ~,
\nonumber \\
&&
= \mu_g \mathbb{S}_g + k_{eg} \mu_e \mathbb{S}_e - ( i \mathbb{F} / v_0 )^\ast \circ \mathbb{F}  ~,
\end{eqnarray}
where $\mu$ , $\mu_g$ , and $\mu_e$ are coefficients. $\mu_g < 0$ , and $\mu_e > 0$. $\ast$ denotes the conjugation of octonion. In the case for single one particle, a comparison with the classical field theory reveals that, $\mathbb{S}_g = m \mathbb{V}_g$ , and $\mathbb{S}_e = q \mathbb{V}_e$ . $m$ is the mass density, while $q$ is the electric charge density. For the charged particle, there may be $\mathbb{V}_e = \mathbb{V}_g \circ \textbf{I} ( \emph{\textbf{I}}_j )$. The unit $\textbf{I} ( \emph{\textbf{I}}_j )$ is one function of $\emph{\textbf{I}}_j$ , with $\textbf{I} ( \emph{\textbf{I}}_j )^\ast \circ \textbf{I} ( \emph{\textbf{I}}_j ) = 1$.

The above can be separated into two field equations,
\begin{eqnarray}
&& \mu_e \mathbb{S}_e = - \lozenge^\ast \circ \mathbb{F}_e  ~,
\\
&& \mu_g \mathbb{S}_g = - \lozenge^\ast \circ \mathbb{F}_g  ~,
\end{eqnarray}
where the former is able to deduce the Maxwell's equations in the classical electromagnetic theory, while the latter is capable of inferring the gravitational inverse-square law in the classical gravitational theory when $\textbf{a} = 0$ and $\textbf{b} = 0$ (see Ref.[19]).

\begin{table}[h]
\caption{The multiplication of the operator with the physics quantity in the complex octonion space.}
\center
{\begin{tabular}{ll}
\hline\hline
definition                         &   expression                                                                 \\
\hline
$\nabla \cdot \textbf{a}$          &  $-(\partial_1 a_1 + \partial_2 a_2 + \partial_3 a_3)$                       \\
$\nabla \times \textbf{a}$         &  $\emph{\textbf{i}}_1 ( \partial_2 a_3
                                      - \partial_3 a_2 ) + \emph{\textbf{i}}_2 ( \partial_3 a_1
                                      - \partial_1 a_3 ) + \emph{\textbf{i}}_3 ( \partial_1 a_2
                                      - \partial_2 a_1 )$                                                         \\
$\nabla a_0$                       &  $\emph{\textbf{i}}_1 \partial_1 a_0
                                      + \emph{\textbf{i}}_2 \partial_2 a_0
                                      + \emph{\textbf{i}}_3 \partial_3 a_0  $                                     \\
$\partial_0 \textbf{a}$            &  $\emph{\textbf{i}}_1 \partial_0 a_1
                                      + \emph{\textbf{i}}_2 \partial_0 a_2
                                      + \emph{\textbf{i}}_3 \partial_0 a_3  $                                     \\

$\nabla \cdot \textbf{A}$          &  $-(\partial_1 A_1 + \partial_2 A_2 + \partial_3 A_3) \emph{\textbf{I}}_0 $  \\
$\nabla \times \textbf{A}$         &  $-\emph{\textbf{I}}_1 ( \partial_2
                                      A_3 - \partial_3 A_2 ) - \emph{\textbf{I}}_2 ( \partial_3 A_1
                                      - \partial_1 A_3 ) - \emph{\textbf{I}}_3 ( \partial_1 A_2
                                      - \partial_2 A_1 )$                                                         \\
$\nabla \circ \textbf{A}_0$        &  $\emph{\textbf{I}}_1 \partial_1 A_0
                                      + \emph{\textbf{I}}_2 \partial_2 A_0
                                      + \emph{\textbf{I}}_3 \partial_3 A_0  $                                     \\
$\partial_0 \textbf{A}$            &  $\emph{\textbf{I}}_1 \partial_0 A_1
                                      + \emph{\textbf{I}}_2 \partial_0 A_2
                                      + \emph{\textbf{I}}_3 \partial_0 A_3  $                                     \\
\hline\hline
\end{tabular}}
\end{table}

\section{Force equilibrium equation}

In the octonion space, from the octonion field source $\mathbb{S}$ , the octonion linear momentum density $\mathbb{P}$ can be defined as,
\begin{eqnarray}
\mathbb{P} = \mu \mathbb{S} / \mu_g   ~,
\end{eqnarray}
where $\mathbb{P} = \mathbb{P}_g + k_{eg} \mathbb{P}_e$ . The octonion linear momentum density consists of two components. One component is, $\mathbb{P}_g = \mathbb{S}_g - i \mathbb{F}^\ast \circ \mathbb{F} / ( v_0 \mu_g )$ , in the quaternion space $\mathbb{H}_g$ . And the other component is, $\mathbb{P}_e = \mu_e \mathbb{S}_e / \mu_g$, in the $S$-quaternion space $\mathbb{H}_e$ .

Further one can define the octonion angular momentum density,
\begin{eqnarray}
\mathbb{L} = ( \mathbb{R} + k_{rx} \mathbb{X} )^\star \circ \mathbb{P}  ~,
\end{eqnarray}
with $\mathbb{L} = \mathbb{L}_g + k_{eg} \mathbb{L}_e$ . And it is able to define the octonion torque density,
\begin{eqnarray}
\mathbb{W} = - v_0 ( i \mathbb{F} / v_0 + \lozenge ) \circ \mathbb{L}  ~,
\end{eqnarray}
where $\mathbb{W} = \mathbb{W}_g + k_{eg} \mathbb{W}_e$ . $\mathbb{L}_g$ and $\mathbb{W}_g$ are components in the quaternion space $\mathbb{H}_g$, while $\mathbb{L}_e$ and $\mathbb{W}_e$ are components in the $S$-quaternion space $\mathbb{H}_e$ . $\mathbb{X}$ is the integrating function of field potential, with $\mathbb{A} = i \lozenge^\star \circ \mathbb{X} $ . $\star$ is the complex conjugate. $k_{rx}$ is a coefficient.

From the octonion torque $\mathbb{W}$, the octonion force density $\mathbb{N}$ is defined as,
\begin{equation}
\mathbb{N} = - ( \lozenge + i \mathbb{F} / v_0 ) \circ \mathbb{W}  ~,
\end{equation}
where $\mathbb{N} = \mathbb{N}_g + k_{eg} \mathbb{N}_e$ . In the quaternion space $\mathbb{H}_g$ , the component of force density is, $\mathbb{N}_g = - ( i \mathbb{F}_g \circ \mathbb{W}_g / v_0 + i k_{eg}^2 \mathbb{F}_e \circ \mathbb{W}_e / v_0 + \lozenge \circ \mathbb{W}_g )$, which is able to deduce the force, precessional angular velocity, mass continuity equation (Table 2) and so forth. In the $S$-quaternion space $\mathbb{H}_e$ , the component of force density is, $\mathbb{N}_e = - ( i \mathbb{F}_g \circ \mathbb{W}_e / v_0 + i k_{eg}^2 \mathbb{F}_e \circ \mathbb{W}_g / v_0 + \lozenge \circ \mathbb{W}_e )$, which is capable of inferring the current continuity equation and so forth (see Ref.[20]).

In the quaternion space $\mathbb{H}_g$ , the component $\mathbb{N}_g$ of octonion force density can be expanded into
\begin{equation}
\mathbb{N}_g = i N_{10}^i + N_{10} + i \textbf{N}_1^i + \textbf{N}_1  ~,
\end{equation}
where $N_{10}^i$ is the torque divergence. $N_{10}$ is the power density. $\textbf{N}_1^i$ is the force density. $\textbf{N}_1$ is the torque derivative. $\textbf{N}_1^i = \Sigma N_{1k}^i \emph{\textbf{i}}_k$ . $\textbf{N}_1 = \Sigma N_{1k} \emph{\textbf{i}}_k$ . $N_{1j}$ and $N_{1k}^i$ are all real.

When $\mathbb{N}_g = 0$, the force equilibrium equation is able to be derived from $\textbf{N}_1^i = 0$, including the gravity, inertial force, Lorentz force, energy gradient and so forth in the classical field theory.

From the force equilibrium equation, $\textbf{N}_1^i = 0$, it is able to obtain approximately,
\begin{eqnarray}
&& - \partial_0 ( \textbf{p} v_0) + \{ p_0 + W_E / (k_p v_0) \} \textbf{g} / v_0 - \textbf{b} \times \textbf{p}
\nonumber
\\
&& ~~~~~~~
+ k_{eg}^2 ( \textbf{E} \circ \textbf{P}_0 / v_0 - \textbf{B} \times \textbf{P} ) - \nabla (p_0 v_0 + W_E / k_p ) = 0  ~,
\end{eqnarray}
where $\{ p_0 + W_E / (k_p v_0) \} \textbf{g} / v_0$ is the gravity density, $ \partial_0 ( - \textbf{p} v_0 )$ is the inertial force density, $k_{eg}^2 ( \textbf{E} \circ \textbf{P}_0 / v_0 - \textbf{B} \times \textbf{P} )$ is the density of electromagnetic force. $- \nabla (p_0 v_0 + W_E / k_p )$ is the gradient of energy density. $W_E = - \{ ( p_0 a_0 + \textbf{a} \cdot \textbf{p}) + k_{eg}^2 ( \textbf{A}_0 \circ \textbf{P}_0 + \textbf{A} \cdot \textbf{P} ) \}$, is the sum of the gravitational and electromagnetic potential energy. $k_p = (k - 1)$ is the coefficient, with $k$ being the dimension of vector, $\textbf{r} = \Sigma r_k \emph{\textbf{i}}_k$. $\mathbb{P}_g = p_0 + \textbf{p} $, with $\textbf{p} =  \Sigma p_k \emph{\textbf{i}}_k $ . $\mathbb{P}_e = \textbf{P}_0 + \textbf{P} $, with $\textbf{P} =  \Sigma P_k \emph{\textbf{I}}_k $ and $\textbf{P}_0 =  P_0 \emph{\textbf{I}}_0 $ . Comparing with the electromagnetic force in the classical field theory states that, $k_{eg}^2 = \mu_g / \mu_e < 0$ .

\section{Gradient force}

In the force equilibrium equation, it is able to define the inertial mass $m$ from the inertial force term, ${- \partial_0 ( \textbf{p} v_0)}$. And it is capable of defining the gravitational mass, $m_g = \{ p_0 + W_E / (k_p v_0) \} / v_0$ , from the gravity term, $\{ p_0 + W_E / (k_p v_0) \} \textbf{g} / v_0$. The gravitational mass consists of three parts: a) the inertial mass, $m$ ; b) the mass variable term, $m^\prime = - \mathbb{F}^\ast \circ \mathbb{F} / ( v_0^2 \mu_g )$ , caused by the norm of field strength; c) the mass variable term, $m^{\prime\prime} = { W_E / (k_p v_0^2 )}$ , caused by the potential energy. That is,
\begin{eqnarray}
m_g = m + m^\prime + m^{\prime\prime}    ~,
\end{eqnarray}
where the mass term, caused by the potential energy or the norm of field strength, will be converted into one fraction of gravitational mass, in the strong electromagnetic field. And it leads to the gravitational mass to depart from the inertial mass to a certain extent. Further the gravitational mass $m_g$ will be varied with the fluctuation of field strength and of potential energy. Only in the situation when the field strength and potential energy both can be neglected, the mass term $m^\prime$ and $m^{\prime\prime}$ both will be approximate to zero, the gravitational mass $m_g$ may be very close to the inertial mass $m$ at the time.

In the force equilibrium equation, the energy gradient is one component of force. Sometimes the energy gradient is called as the gradient force also, and that it contains a few terms obviously. In the above the gradient force is,
\begin{eqnarray}
\textbf{N}_B = - \nabla (p_0 v_0 + W_E / k_p )  ~,
\end{eqnarray}
where $W_E = ( b^2 - g^2 / v_0^2 ) / ( 2 \mu_g ) + ( B^2 + E^2 / v_0^2 ) / ( 2 \mu_e )$ .

When $ \nabla m = 0$ and $k = 3$, the above can be reduced to
\begin{eqnarray}
&& \textbf{N}_B = \nabla ( b^2 - g^2 / v_0^2 ) / \mu_g - \nabla ( b^2 - g^2 / v_0^2 ) / ( 2 \mu_g )
\nonumber
\\
&& ~~~~~~~
+ \nabla ( B^2 - E^2 / v_0^2 ) / \mu_e - \nabla ( B^2 + E^2 / v_0^2 ) / ( 2 \mu_e ) ~ ,
\end{eqnarray}
where $b^2 = \textbf{b}^\ast \cdot \textbf{b}$, $g^2 = \textbf{g}^\ast \cdot \textbf{g}$. $E^2 = \textbf{E}^\ast \cdot \textbf{E}$, $B^2 = \textbf{B}^\ast \cdot \textbf{B}$. $m^\prime = - ( b^2 - g^2 / v_0^2 ) / ( v_0^2 \mu_g ) - ( B^2 - E^2 / v_0^2 ) / ( v_0^2 \mu_e )$. In order to distinguish from other terms of gradient force, $\textbf{N}_B$ is called as the strength gradient force for the moment in the paper.

When there is only the magnetic flux density $\textbf{B}$ , the above will be reduced to
\begin{eqnarray}
\textbf{N}_B = \nabla B^2 / ( 2 \mu_e )  ~,
\end{eqnarray}
where the magnitude of strength gradient force $\textbf{N}_B$ is in direct proportion to $B$ and $\nabla B$ . However $ \nabla B^2$ is independent to the direction of magnetic flux density $\textbf{B}$ . It means that the strength gradient force $\textbf{N}_B$ will push the charged/neutral particle to move along the gradient direction.

In the force equilibrium equation, the force term contains the gravity, inertial force, Lorentz force, strength gradient force and so forth. The fluctuation of any one force term will impact the state of equilibrium. In other words, even in the situation when the Lorentz force equals to zero, the fluctuation of some force terms has still an influence on the observation result of E\"{o}tv\"{o}s experiment. Especially the fluctuation of electromagnetic strength with the uniform or non-uniform distribution will result in the observable effect of the variable gravitational mass.

The strength gradient force may possess several significant and latent applications in the future. In most cases, the variation of gravitational mass is quite tiny in general, and is very untoward to be measured directly. But the derivative (a force term) of this tiny variation of gravitational mass may generate some considerable physical effects and utilizations.

1) Astrophysical jet. Since a long time, the dynamic principle of astrophysical jets has been puzzling the existing physical theories. In recent years, the strength gradient force is able to be applied to elucidate some dynamic properties of astrophysical jets, in the electromagnetic and gravitational theories, described with the complex octonions. Because the gravitational mass is variable, the relevant strength gradient force, $\textbf{N}_B$ , must be distinct from the strength gradient force, $\textbf{N}_{B(N)}$ , in the Newtonian theory, especially the magnitude and direction. In the complex-octonion electromagnetic and gravitational theories, when the strength gradient force is considered as the thrust of astrophysical jets, the strength gradient force, $\textbf{N}_B$ , which is only relevant to the gravitational strength, will propel various ingredients within the astrophysical jet to launch into the faraway interstellar spaces (see Ref.[17]).

2) Dynamic force. The dynamic principle of astrophysical jets can be extended into other research domains obviously. According to the complex-octonion electromagnetic and gravitational theories, the strength gradient force, $\textbf{N}_B$ , which is relevant to the electromagnetic strength, may be applied to impel multiform mechanisms in the laboratory. For instance, the strength gradient force in the ultra-strong magnetic field can be chosen as the thrust to drive the precision moving mechanism \cite{weng6}. On the other hand, the strength gradient force relates merely to the gradient of the norm of electromagnetic strength, in the electromagnetic fields. That is, the strength gradient force is independent of not only the direction of electromagnetic strength but also the mass and electric charge for the test particles. The strength gradient force is capable of transporting various matters (ordinary matter and dark matter) from one location to another. As a result, the strength gradient force can be utilized to construct the working principle for the all-particle accelerator or molecule decelerator \cite{narevicius,meijer}, and even the propulsion principle \cite{white} of a new vehicle.

\begin{table}[h]
\caption{Some definitions in the gravitational and electromagnetic field in the complex octonion space.}
\center
{\begin{tabular}{ll}
\hline\hline
physics~quantity             &   definition                                                                            \\
\hline
radius~vector                &  $\mathbb{R} = \mathbb{R}_g + k_{eg} \mathbb{R}_e  $                                    \\
integrating~function         &  $\mathbb{X} = \mathbb{X}_g + k_{eg} \mathbb{X}_e  $                                    \\
field~potential              &  $\mathbb{A} = i \lozenge^\star \circ \mathbb{X}  $                                     \\
field~strength               &  $\mathbb{F} = \lozenge \circ \mathbb{A}  $                                             \\
field~source                 &  $\mu \mathbb{S} = - ( i \mathbb{F} / v_0 + \lozenge )^\ast \circ \mathbb{F} $          \\
linear~momentum              &  $\mathbb{P} = \mu \mathbb{S} / \mu_g $                                                 \\
angular~momentum             &  $\mathbb{L} = ( \mathbb{R} + k_{rx} \mathbb{X} )^\star \circ \mathbb{P} $              \\
octonion~torque              &  $\mathbb{W} = - v_0 ( i \mathbb{F} / v_0 + \lozenge ) \circ \mathbb{L} $               \\
octonion~force               &  $\mathbb{N} = - ( i \mathbb{F} / v_0 + \lozenge ) \circ \mathbb{W} $                   \\
\hline\hline
\end{tabular}}
\end{table}

\section{Experiment proposal}

The predictions on the strength gradient force in the Newtonian mechanics are distinct from that in the electromagnetic and gravitational theories, described with the complex octonions (Table 3). The predicting strength gradient force is, $\textbf{N}_{B(N)} = - \nabla ( W_E / k_p )$, in the Newtonian mechanics. And the predicting strength gradient force is, $\textbf{N}_B = - \nabla (p_0 v_0 + W_E / k_p )$ , in the electromagnetic and gravitational theories, described with the complex octonions. It is found that there may be not only the magnitude differentia but also the vector discrepancy between two force terms, $\textbf{N}_{B(N)}$ and $\textbf{N}_B$ , in terms of the predicting strength gradient forces of two theories. By means of certain appropriate experiments, it must be possible to discriminate the contradistinctions between these two theories.

In the complex octonion space, it is able to validate directly the influence of the electromagnetic strength on the gravitational mass $m_g$ or the strength gradient force $\textbf{N}_B$ in the laboratory. For instance, in the strong magnetic field, it is capable of verifying the mass term $m^\prime$ and the strength gradient force $\textbf{N}_B$ , and their influence on the E\"{o}tv\"{o}s experiment.

In the strong magnetic field, it is feasible to validate the E\"{o}tv\"{o}s experiment. According to the force equilibrium equation, $\textbf{N}_1^i = 0$, the gravitational mass is approximately written as, $m_g = m + m^\prime + m^{\prime\prime} $ . However each E\"{o}tv\"{o}s experiment has never been inspected under the electromagnetic environment until now. Consequently it is necessary to validate the E\"{o}tv\"{o}s experiment in the strong magnetic field. a) When the strong magnetic field is one uniform distribution, the variation of magnetic flux density will alter the gravitational mass. b) In case the distribution of strong magnetic field is non-uniform (such as the pulsing magnetic field), the variation of magnetic flux density will result in not only the alteration of gravitational mass but also the emergence of strength gradient force. As a result, the strong magnetic field must disturb the previous state of equilibrium, transferring the equilibrium situation of the neutral particle from one location to another.

The above analysis states that the gravitational mass $m_g$ and strength force $\textbf{N}_B$ have an influence on the E\"{o}tv\"{o}s experiment and other experiments. For the sake of validating effectively the E\"{o}tv\"{o}s experiment in the strong magnetic field, it is beneficial to develop the following auxiliary experiment, in order to confirm to each other.

The neutral particle will be repelled by the electric field with the non-uniform distribution. In the strong electric field, the strength gradient force, $\textbf{N}_B = - 3 \nabla E^2 / ( 2 v_0^2 \mu_e )$, yielded by the electric field intensity $\textbf{E}$ is able to impact the state of equilibrium. The strength gradient force yielded by the electric field intensity is along the reverse direction of gradient direction. a) When the electric field intensity is weak, the electrostatic force (attractive) caused by the electrostatic induction is too weak to attract the neutral particle, while the strength gradient force (repulsive) produced by the electric field intensity is feeble too. b) When the electric field intensity is comparative strong, the electrostatic force caused by the electrostatic induction is strong enough, and stronger than the strength gradient force yielded by the electric field intensity, so the electric field is able to attract the neutral particle. c) When the electric field intensity is very strong, the electrostatic force caused by the electrostatic induction will be weaker than the strength gradient force yielded by the electric field intensity, consequently the electric field is unable to attract the neutral particle. Further the neutral particle will be pushed away by the strong electric field, along the reverse direction of gradient direction.

In the recent experiment conditions, it is comparatively untoward to attempt to measure directly the tiny variation of gravitational mass. The scholars may require a few new experimental devices, besides the existing devices in the E\"{o}tv\"{o}s experiments. Presenting a striking contrast to the above is that it will be much more uncomplicated to measure the energy gradient, which is relevant to the variation of gravitational mass. That is the contrastive study on the magnitude and/or direction between two force terms, $\textbf{N}_B$ and $\textbf{N}_{B(N)}$ . Apparently this is an indirect approach. The paper proposes to meliorate the existing experimental devices in Ref.[34] or Ref.[35]. One more proposal scheme is to amend the existing devices in Ref.[36], enabling the amended devices to be applied in the ordinary (or non-vacuum) conditions. These three sorts of improved experimental devices are capable of measuring the energy gradient. The order of magnitude of the corrections expected of the three proposed experiments may be approximately equivalent to that of their original experiments respectively. Of more important is to arrange effectively the distributions of electric or magnetic fields, configuring the particle time-of-flight in these distributed fields.

In the strong electromagnetic field, the validation of above experiments proposal will be beneficial to investigate the influence of gravitational mass $m_g$ and of strength gradient force $\textbf{N}_B$ on the E\"{o}tv\"{o}s experiment.

\begin{table}[h]
\caption{A Comparison of the energy gradient between the Newtonian theory and the octonion electromagnetic and gravitational field theory, in the E\"{o}tv\"{o}s experiments.}
\center
{\begin{tabular}{llll}
\hline\hline
field~theory                   &   energy~gradient                                        &    gravitational~mass      &    inertial~mass     \\
\hline
Newtonian~theory               &   $\textbf{N}_{B(N)} = - \nabla ( W_E / k_p )$           &    invariable              &    invariable        \\
octonion~field~theory          &   $\textbf{N}_B = - \nabla ( p_0 v_0 + W_E / k_p )$      &    variable                &    invariable        \\
\hline\hline
\end{tabular}}
\end{table}

\section{Discussions and conclusions}

Ever since a long time ago, the scholars have a suspicion of that the validity of E\"{o}tv\"{o}s experiment may not be effective forever. The scholars are being beleaguered by this perplexing puzzle until now. However the existing theory is at a loss to explain why the gravitational mass seems to be equal to the inertial mass in the most cases. What the most important thing is that most of existing E\"{o}tv\"{o}s experiments are merely measured in the weak gravitational field on the surface of the earth until now. And it has never been surveyed in the strong gravitational fields. Especially the E\"{o}tv\"{o}s experiment has not been tested in the comparatively strong electromagnetic field either. Obviously it is necessary to consider more influencing factors in the E\"{o}tv\"{o}s experiment, including the strong electromagnetic strength with the uniform or non-uniform distributions.

In the electromagnetic and gravitational theory described with the complex octonion, it is able to deduce the force equilibrium equation, variable gravitational mass, strength gradient force and so forth. In the complex octonion space, the variation of gravitational mass is relevant to the electromagnetic and gravitational strength. The gradient of norm of the electromagnetic and gravitational strength produces the strength gradient force. This force term is dependent on neither the direction of field strength, nor the mass and electric charge of charged particles. This property will impact the gravitational mass and the force equilibrium equation in the E\"{o}tv\"{o}s experiment. Meanwhile it has an influence on other experiments also, including the experiment about the electric field with the gradient distribution impelling the neutral particles.

It should be noted that the paper investigated only some simple properties of the strength gradient force and the variable gravitational mass in the electromagnetic and gravitational field. Despite its preliminary characteristics, this study can clearly indicate that some influencing factors are able to impact the E\"{o}tv\"{o}s experiment in the strong magnetic field. And it lays the foundation for continuing researches of the property of gravitational mass in the strong magnetic field. In the following study, it is going to explore the E\"{o}tv\"{o}s experiment in the strong pulsing magnetic field, and validate the repelling influence of the electric field with the gradient distribution on the neutral particles in the strong electric field.

\begin{acknowledgements}
The author is indebted to the anonymous referees for their valuable comments on the previous manuscripts. This project was supported partially by the National Natural Science Foundation of China under grant number 60677039.
\end{acknowledgements}



\end{document}